# Causal impact of severe events on electricity demand: The case of COVID-19 in Japan

Yasunobu Wakashiro [1]

*Center for Social Systems Innovation, Kobe University*

**Abstract**

As of May 2022, the coronavirus disease 2019 (COVID-19) still has a severe global impact on people's lives. Previous studies have reported that COVID-19 decreased the electricity demand in early 2020. However, our study found that the electricity demand increased in summer and winter even when the infection was widespread. The fact that the event has continued over two years suggests that it is essential to introduce the method which can estimate the impact of the event for long period considering seasonal fluctuations. We employed the Bayesian structural time-series model to estimate the causal impact of COVID-19 on electricity demand in Japan. The results indicate that behavioral restrictions due to COVID-19 decreased the daily electricity demand (-5.1% in weekdays, -6.1% in holidays) in April and May 2020 as indicated by previous studies. However, even in 2020, the results show that the demand increases in the hot summer and cold winter (the increasing rate is +14% in the period from 1st August to 15th September 2020, and +7.6% from 16th December 2020 to 15th January 2021). This study shows that the significant decrease in electricity demand for the business sector exceeded the increase in demand for the household sector in April and May 2020; however, the increase in demand for the households exceeded the decrease in demand for the business in hot summer and cold winter periods. Our result also implies that it is possible to run out of electricity when people's behavior changes even if they are less active.

## 1. Introduction

COVID-19 which was found firstly in 2019 has a severe influence on economic activities globally still in May 2022. In many countries and regions, the rapid spread of the infection has had an impact on people's health, education, business and everyday life. Facing the rapid infection, many governments restricted the movement of people to implement lockdown which forced people not to go out. In Japan, the government just announced a state of emergency to ask people to stay home instead of lockdown, because it is not understood that the Japanese legal system allows to implement lockdown. The declaration was announced two times by the end of March 2021, the first is declared from April 7th to May 26th in 2020 and the second is from January 1st to March 7th (Table 1). During the first emergency period, the government asked restaurants, bars, stores, gyms, and companies to close as well as restricted people from moving.

[*] Corresponding author. Kobe-city, Hyogo-ken, Japan.
*E-mail address:* yasunobu.wakashiro@gmail.com

**Table 1.** Declaration of Emergency in Japan by the end of March 2021

| Periods | Prefectures and areas in a state of emergency |
| --- | --- |
| April 7th– April 16th | 7 prefectures in 3 areas |
| April 8th–May 6th | All prefectures |
| May 7th–May 14th | 13 prefectures in 6 areas |
| May 15th–May 21st | 8 prefectures in 3 areas |
| May 22nd–May 25th | 4 prefectures in 2 areas |
| May 26th– January 7th | All prefectures are off a state of emergency |
| January 8th– March 21st | 4 prefectures in 1 area |
| January 14th– March 7th | 7 prefectures in 3 area |

Japan has experienced three peaks of infection from February 2020 to March 2021. The first peak is around the latter of April 2020, the second is in August 2020, and the third is in the middle of January 2021 (Figure 1).

**Figure 1.** The number of the infected in Japan

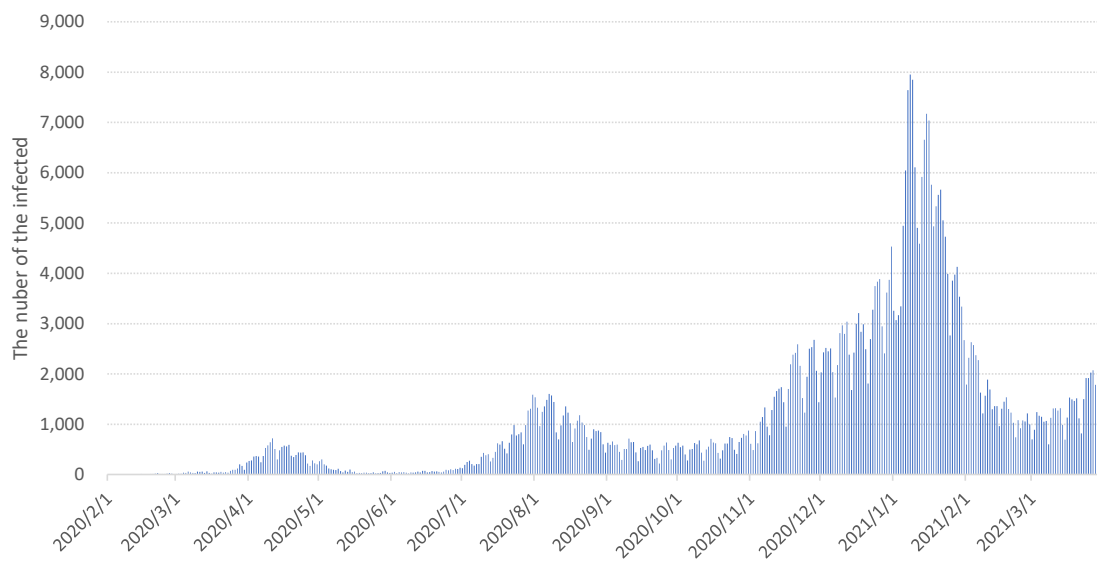

Many studies have reported that the demand for electricity decreased globally as an impact of the COVID-19 pandemic in early and mid-2020. Elavarasan et al. (2020) studied the demand of power in the USA, Italy, Australia, India, and other countries to find that COVID-19 has changed individuals' lifestyle globally, since people are primarily restricted to their homes. Carvalho et al. (2021) pointed out that the electricity consumption patterns and levels in Brazil have changed as a result of the COVID-19 crisis. A decrease in the electricity consumption levels were observed that were found to be statistically significant. García et al. (2021) showed that residential consumers increased their electricity consumption during the lockdown and reopening periods. In contrast, the global non-residential consumers decreased their electricity consumption during the lockdown and reopening periods. Kirli et al. (2021) analyzed the impact of the implementation of a lockdown due to the COVID-19 crisis on the electricity system of Great Britain and concluded that the lockdown not only decreased the demand for electricity, but also changed its pattern of consumption after a majority of the commercial users shut down. In the case of Japan, Wakashiro (2020) estimated the impact of the state of emergency on the electricity demand from March 2020 to May 2020 and concluded that the decrease in the electricity demand could be attributed to the decline in the activities of the industrial and business sectors.

**Table 2.** Change in the demand for electricity observed in existing studies

| Articles | Country, Area | Period | Change of rate |
|---|---|---|---|
| Elavarasan et al. (2020) | New York, USA | Lockdown periods in each region or countries | -2.9% in February to -13.7% in April |
|  | Australia |  | -6.7% in March |
|  | France |  | -14.80% |
|  | Italy |  | -11% |
|  | Spain |  | -7.10% |
|  | Portugal |  | -6.40% |
|  | Belgium |  | -6.10% |
|  | Netherland |  | -5% |
|  | Germany |  | -4.90% |
|  | UK |  | -6% |
| Bahmanyar et al. (2020) | Spain | The second week of April 20202 | -25% |
|  | Italy |  | -17.70% |
|  | Belgium |  | -15.60% |
|  | UK |  | -14.20% |
|  | Netherland |  | -11.60% |
|  | Sweden |  | 2.10% |
| Carvalho et al. (2020) | Brazil | January 1 and May 27, 2020 | -20% (The Southeast-Midwest area) |
| Alasali et al. (2020) | Jordan | The periods during the movement restriction orders | -33% (Maximum half-hourly demand) |
| Abu-Rayash and Dincer (2021) | Ontario | April | -14% |
| Agdas and Barooah (2020) | North Central Florida | May | +10% |

As seen above, many studies have analyzed the impact of lockdowns on electricity consumption during the COVID-19 crisis from various perspectives and concluded that they have decreased the electricity demand. However, whether lockdowns have always decreased the electricity demand, even when its demand for households increased, remains unclear. Table 2 shows the changes of electricity demand in the previous studies.

This paper is structured as follows. Section 2 constructs the Bayesian structural time-series model and partial adjustment model, and also explains the data used of analyses. Section 3 presents the results of the analyses. In Section 4, we discuss the implication of the result. Finally, Section 5 concludes the study.

## 2. Methods and Materials

In this section, we describe the method of estimating the causal impact on electricity demand, construct the model to analyze the factor affecting the impact, and also present extracting procedure for the data used for analyses.

### *2.1. Constructing Bayesian structural time-series model*

As described in the previous section, some studies have simply compared the electricity demand in the period of COVID-19 and the demand in the period before the disease. However, it is desirable to infer counterfactual data series which accurately reflect seasonal and weekly fluctuations because the simply aggregated data may include outliers made by temperature or day-specific factors. We employ the Bayesian structural time-series model which is originally implemented by Brodersen et al. (2015) (this model is employed in energy economics literatures, for example Märkle-Huß et al. (2018)). This model is originally developed as a means of estimating the impact of marketing trials for the long run, and it constructs the most appropriate synthetic control for modelling the counterfactual of a time series observed both before and after the event.

We can estimate the impact of the COVI-19 on electricity demand by subtracting the actual demand from the counterfactual demand inferred as below[2]. The causal impact of COVID-19 is defined as

$$y_t - \tilde{y}_t,$$

where $y_t$ is actual observed electricity demand, and $\tilde{y}_t$ is counterfactual demand, both after the event. The structural time-series model which we employ here, can be define as a pair of the observed equation which links the observed data $y_t$ to a state vector $\alpha_t$ and state equation which governs the evolution of the state vector $\alpha_t$ through time as below[3].

$$y_t = Z_t^T \alpha_t + \varepsilon_t,$$
$$\alpha_{t+1} = T_t \alpha_t + R_t \eta_t,$$

where $Z_t$ is an output vector, $T_t$ is a transition matrix, $R_t$ is a control matrix, $\varepsilon_t$ is a scalar observation error with noise variance $\sigma_t$, and $\eta_t$ is a system error.

---

[2] Readers who need to know more detailed discussion, see Brodersen et al. (2015)

[3] $\varepsilon_t \sim \mathcal{N}(0, \sigma_t^2)$ and $\eta_t \sim \mathcal{N}(0, Q_t)$ are assumed to be independent of all other unknowns.

Accurate counterfactual prediction needs covariates to explain unobserved factors which the model does not account for. We assume a static coefficients linear regression model to include the control series in the model, so a static regression can be written in state-space form by setting $Z_t = \beta^T x_t$, and all covariates are assumed to be contemporaneous. The covariates should be unaffected by the event, so we define the time-series before the infection widespread as the covariate data. The inferring period is defined as $t = 1, I, m$, and is supposed that the event occurs at $t = n + 1$. The prior time-series before the event is $1, I, n$, and posterior time-series is $n + 1, \ldots, m$.

The counterfactual time series electricity demand $\tilde{y}_t$ is constructed as follows. The set of all model parameters $\theta$ and the state vector $\alpha = (\alpha_1, \ldots, \alpha_m)$ are inferred by specifying a prior distribution. Inference is conducted by Bayesian approach using MCMC, where a prior distribution for such a variance follows the Gamma distribution. Posterior inference in our model consists of three pieces. First, we simulate draws of the model parameters $\theta$ and the state vector $\alpha$ given the observed data $y_{1:n}$ in the training period. Second, we use the posterior simulations to simulate from the posterior predictive distribution $p(\tilde{y}_{n+1:m}|y_{1:n})$ over the counterfactual time series $\tilde{y}_{n+1:m}$ given the observed pre-intervention activity $y_{1:n}$. Third, we use the posterior predictive samples to compute the posterior over the counterfactual time series $\tilde{y}_{n+1:m}$ given the observed pre-intervention ac- distribution of the pointwise impact $y_t - \tilde{y}_t$ for each $t = 1, \ldots, m$. We use the same samples to obtain the posterior distribution of cumulative impact. Gibbs sampler is used to simulate a sequence $(\theta, \alpha)^{(1)}, (\theta, \alpha)^{(2)}, \ldots$ from a Markov chain whose stationary distribution is $p(\theta, \alpha|y_{1:n})$.

*2.2. Employing partial adjustment model using GMM estimator*

The formulation in the previous subsection allows us to obtain the difference of the actual demand and counterfactual data. To examine which factor affect the difference, we employed the partial adjustment model using the generalized method of moments referring to Wakashiro (2019).

The estimation model is formulated as below.

$$\text{Ln } ELE_{area,t} = \alpha + \beta_1 \ln Resi_{area,t} + \beta_2 \ln Wrk_{area,t} + \beta_3 \ln Retl_{area,t} + \beta_3 \ln Grcy_{area,t} + \beta_4 \ln ELE_{area,t-1} + \Delta\mu$$

where ln represents the natural logarithm, area (area = 1, 2, …, 10) stands for the area which is explained in the next subsection, and t (t = 1, 2, …, T) means dates. The dependent variable, $ELE_{area,t}$, is the change rate of electricity between the actual and counterfactual electricity demand. Independent variables are defined as follows. $Resi_{area,t}$, $Wrk_{area,t}$, $Retl_{area,t}$, $Grcy_{area,t}$ are the change rates of the number of people who stay in home, workplace, retail and recreation place, and grocery and pharmacy, respectively from the previous year. One of the independent variables, $ElE_{area,t-1}$, is the change rate of electricity demand in the previous day, which indicates that electricity demand depends not only on electricity use in the present day but also on use in the previous day. Tanishita (2009) used a partial adjustment model based on OLS (ordinary least square) estimation. However, the paper indicates that a lagged dependent variable has the possibility of endogeneity (Hsiao, 2002). In other words, a lagged dependent variable correlates with the error term. This dynamic panel bias would make the estimated long-run price elasticity higher than the true value. The electricity demand is affected by other factors beyond those captured by the independent variables, as seen in the relationship between the electricity price and the error term. To avoid these biases, we employ an additional lag of a lagged dependent variable, $ELE_{area,t-2}$ as instrumental variables.

*2.3. Extracting daily and hourly electricity demand for each area*

Many previous studies straightforwardly use the electricity demand data to compare the period of the lockdown and the pre-lockdown period. Alasali et al. (2021) compared electricity demand in 2020 to the demand between 2016-2019. Kirli et al. (2021) compared the electricity demand in lockdown week to the demand in a pre-lockdown week. Bahmanyar et al. (2020) compared the hourly load profile comparison for the second week of April 2020 and a reference week in 2019 for Peninsular Spain, Italy, the UK, Belgium, Netherlands, and Sweden. Ruan et al. (2021) used open-access data which was established to track the impacts of COVID-19 on 7 electricity markets in the U.S., and also showed the electricity consumption from February to mid-July using the last three years data which are properly aligned to allow comparisons between the same weekdays. Abu-Rayash and Dincer (2020) simply compared hourly electricity demand all of the day in April 2019 and in April 2020, and also compared the demand of the same day in April 2019 and April 2020. Buechler et al. (2020) compared between April 2020 daily load curves and historic load curve patterns (April 2016, 2017, 2018 and 2019). Wakashiro (2020) analyzed the daily electricity demand by adjusting the temperature on holidays and weekdays. They estimated the adjusted electricity demand in Japan from March to May in 2020. Santiago et al. (2021) measured daily power demand profiles, especially morning and evening peaks, have been modified at homes, hospitals, and in the total power demand.

We use daily electricity demand from April 1$^{st}$ 2016 to March 31$^{st}$ 2021. We obtained electricity demand data which contain the 24-hour data for 365 days from the electricity distribution companies' websites in ten areas (Hokkaido Electric Power Network, Inc. 2021; Tohoku Electric Power Company, Inc. 2021; TEPCO Power Grid 2021; Chubu Electric Power Grid Company, Inc. 2021; Hokuriku Electric Power Transmission & Distribution Company 2021; Kansai Transmission and Distribution, Inc. 2021; Chugoku Electric Power Transmission & Distribution Company, Inc. 2021; Shikoku Electric Power Transmission & Distribution Company, Inc. 2021; Kyushu Electric Power Transmission and Distribution Company, Inc. 2021; The Okinawa Electric Power Company, Inc. 2021). Each area covers the prefectures as shown in Table 3. We should note that the daily and hourly electricity demand before April 2016 was not published by electric distribution company in each area.

**Table 3.** Ten areas and the prefectures

| Areas | Prefectures | Population Density (population/km$^2$, 2019) |
|---|---|---|
| 1. Hokkaido area | Hokkaido | 63.77 |
| 2. Tohoku area | Aomori, Iwate, Miyagi, Akita, Yamagata, Fukushima and Niigata | 149.60 |
| 3. Tokyo area | Tokyo, Kanagawa, Saitama, Chiba, Tochigi, Ibaragi, Gunma, Yamanashi, and Shizuoka | 1,091.35 |
| 4. Chubu area | Aichi, Nagano, Gifu, Mie, and Shizuoka | 397.23 |
| 5. Hokuriku area | Toyama, Ishikawa, and Fukui | 306.79 |
| 6. Kansai area | Shiga, Kyoto, Osaka, Hyogo, Nara, and Wakayama | 761.27 |
| 7. Chugoku area | Hiroshima, Yamaguchi, Shimane, Tottori, and Okayama | 231.59 |
| 8. Shikoku area | Kagawa, Tokushima, Ehime, and Kochi | 201.61 |
| 9. Kyushu area | Fukuoka, Nagasaki, Oita, Saga, Miyazaki, Kumamoto, and Kagoshima | 325.89 |
| 10. Okinawa area | Okinawa | 632.58 |

Source: National census of Japan

*2.4. Manipulating demand for making the actual, covariates, and learning data*

As stated before, covariates and training period must not have correlation to the influence of event. COVID-19 started to affect the severe impact on social activities in Japan from April 2020, the period from April 2019 to March 2020 is used as training sample to estimate accurately. Covariates are defined using the data from April 2016 to March 2019. The actual demand affected by COVID-19 is defined as the period from April 2020 to March 2021.

We must carefully choose the dates of counterfactual data precisely corresponding to the real data. We extracted the days corresponding to the days after the event considering day of the week, since activities on Mondays must be different from those on Fridays. To compare electricity demand on the same day of the week, we make a group of days to compare with the following procedure. We set the date of the first Monday of 2020 (January 6) as the reference date. In each year to be compared, the Monday closest to the reference date was set as the reference date for that year, and the same number was assigned from the reference date of each year. We implemented an accurate daily comparison based on the day of the week by comparing the days with the same number. Comparing demand of each date is performed only if each date is all weekdays or all holidays in the target year. If the days are not all holidays nor weekdays, the day is not analyzed. We should also note that this study uses only weekday electricity demand data. The example of this procedure is shown in Table 4.

**Table4.** Procedure of making a group of days to compare

|     | 2017   |   | 2018   |   | 2019   |   | 2020   |   | Analysis target | Sample No. |
| --- | ------ | - | ------ | - | ------ | - | ------ | - | --------------- | ---------- |
| Mon | 9-Jan  | H | 8-Jan  | H | 7-Jan  |   | 6-Jan  |   | No              | -          |
| Tue | 10-Jan |   | 9-Jan  |   | 8-Jan  |   | 7-Jan  |   | Yes             | 1          |
| Wed | 11-Jan |   | 10-Jan |   | 9-Jan  |   | 8-Jan  |   | Yes             | 2          |
| Thu | 12-Jan |   | 11-Jan |   | 10-Jan |   | 9-Jan  |   | Yes             | 3          |
| Fri | 13-Jan |   | 12-Jan |   | 11-Jan |   | 10-Jan |   | Yes             | 4          |
| Sat | 14-Jan | W | 13-Jan | W | 12-Jan | W | 11-Jan | W | Yes             | 5          |
| Sun | 15-Jan | W | 14-Jan | W | 13-Jan | W | 12-Jan | W | Yes             | 6          |
| Mon | 16-Jan |   | 15-Jan |   | 14-Jan | H | 13-Jan | H | No              | -          |
| Tue | 17-Jan |   | 16-Jan |   | 15-Jan |   | 14-Jan |   | Yes             | 7          |
| Wed | 18-Jan |   | 17-Jan |   | 16-Jan |   | 15-Jan |   | Yes             | 8          |
| Thu | 19-Jan |   | 18-Jan |   | 17-Jan |   | 16-Jan |   | Yes             | 9          |
| Fri | 20-Jan |   | 19-Jan |   | 18-Jan |   | 17-Jan |   | Yes             | 10         |
| Sat | 21-Jan | W | 20-Jan | W | 19-Jan | W | 18-Jan | W | Yes             | 11         |
| Sun | 22-Jan | W | 21-Jan | W | 20-Jan | W | 19-Jan | W | Yes             | 12         |

H…holidays, W…weekends

*2.5. Extracting people's behavior for analyzing the reason of change of electricity demand*

We extracted the change rates of the number of people who stay in home, workplace, retail and recreation place, and grocery and pharmacy, respectively in ten areas as independent variables for analyzing partial adjustment model from the mobility report (Google, 2021). This report shows movement trends by region, across different categories of places, and is created with aggregated, anonymized sets of data from users who have turned on the Location History setting.

## 3. Results

We present the causal impact of COVID-19 on the electricity demand from April 2020 to March 2021. As the readers of this journal might know, the electricity demand in weekdays is different from the one in holidays.

*3.1. Causal impact of the event on all day electricity demand*

Figure 3 shows the actual and counterfactual electricity demand in the period from April 2019 to March 2021. The vertical axis means electricity demand on each day, and the horizontal axis depicts the days from 1$^{st}$ April 2019. The vertical dotted lines drawn in each graph mean the dates that the event occurs (7$^{th}$ April 2020), and the shaded area means 95% confidence intervals. The upper plot describes the counterfactual demand as dashed line, and actual series as the real line. The middle plot shows the difference between the counterfactual demand minus actual time-series, and cumulative sum of the difference is plotted in the bottom graph.

On weekdays, Figure 2 shows a rapid decrease in electricity demand from May to July 2020, the actual demand is basically below the counterfactual line, the actual demand is seen below the confidence interval temporary. Although the demand increases in summer and winter, the cumulative difference shows that the demand decreases on weekdays.

Figure 2 presents the actual demand decreases in the period from spring to early summer, but keenly increases in the hot summer and cold winter. We define the periods to statistically measure the causal impact of the event, and also measure the accumulate change of electricity demand in each period.

**Figure 2.** The causal impact of COVID-19 on electricity demand on weekdays

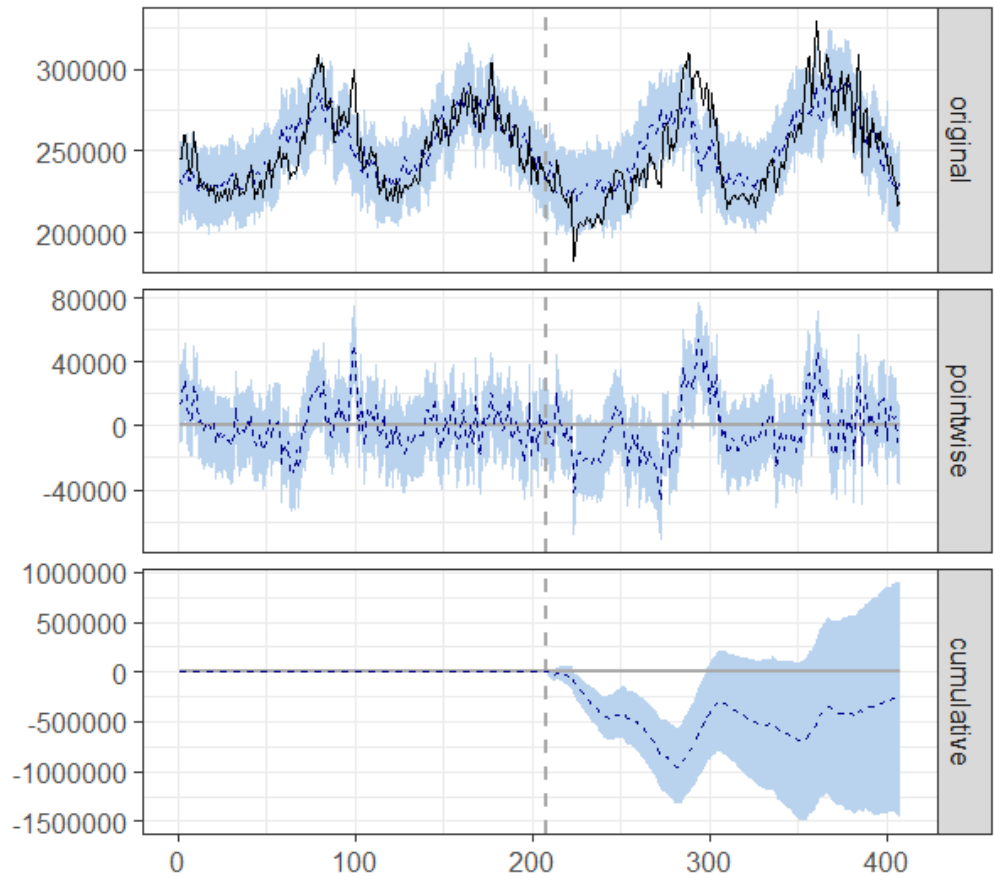

*3.2. Causal impact in each period*

Four seasons are enjoyed in Japan, but the variation of temperature is divided in 5 periods. First of all, the moderate climate is experienced in the period from the spring to early summer (from April to July). People enjoy the moderate climate, and spend their time without air conditioning devices. Secondly, hot summer comes in the period from August to early September, the temperature sometimes reaches near 40 degree Celsius. Many people cannot spend their time without air conditioning devices. Thirdly, the moderate season comes again. In the period from the early September to December, many people can spend their time without devices. Fourthly, the coldest time in Japan comes in the period from later December to early January. Although global warming comes around, air conditioning devices are essential. The period from later January to Mach experiences cold time, but essentiality of the device is weaker. We divided in the five periods. Levene's test shows that the variance of temperature is not homogenous of between periods, and more, Tukey test indicates that the multiple comparisons of means are different between the periods. The results are presented in Table A1 in Appendix.

The result Is presente' table 5. We analyze the electricity demand during the period from April $1^{st}$ to July $31^{st}$ in this period, electricity demand decreased significantly, reaching -5.1% compared to counterfactual demand, the posterior probability is over 99%. From August to the first half of September, the demand increased sharply to 14%. The readers notice that the electricity demand declined from late September to early December, but we should not that this change is not statistically significant at the 95% level. From the latter half of December to the first half of January, it increased again to 7.6%. It is statistically significant at the 95% level. We also note that the change in the period from the late of January to March has decreased to -9.5%, it is not statistically significant as in the autumn.

**Table 5.** Results of inferring the causal impact for each period on weekdays

| From | Till | Relative Effect | Probability |
|---|---|---|---|
| $1^{st}$ Apr | $31^{st}$ Jul | -5.1% | 99.89% |
| $1^{st}$ Aug | $15^{th}$ Sep | 14% | 99.57% |
| $16^{th}$ Sep | $15^{th}$ Dec | -9.6% | 92.34% |
| $16^{th}$ Dec | $15^{th}$ Jan | 7.60% | 97.75% |
| $16^{th}$ Jan | 31th Mar | -9.5% | 93.62% |

*3.3. Sensitivity analysis of inferring the counterfactual demand*

We tested the sensitivity of our result by supposing if the event occurs a year earlier than the actual event. Instead of the day when the event actually happened in 2020, we inferred the counterfactual demand in 2019. Table 6 shows the results. All of the change of electricity demand in the case of the event occurring in 2019 are not statistically significant on weekdays. On holidays, the change in the winter is statistically significant, but the direction of the change differs from the expectation.

**Table 6.** Robustness check

| Periods | | Weekdays | | Holidays | |
| --- | --- | --- | --- | --- | --- |
| From | Till | Relative Effect | Probability | Relative Effect | Probability |
| 1st Apr | 31st Jul | -5.0% | 92.86% | -0.73% | 66.90% |
| 1st Aug | 15th Sep | -6.4% | 92.28% | 0.87% | 69.00% |
| 16th Sep | 15th Dec | -7.7% | 86.14% | -0.24% | 57.80% |
| 16th Dec | 15th Jan | 3.10% | 79.95% | -7% | 99.80% |
| 16th Jan | 31th Mar | -6.6% | 85.12% | -1.7% | 85.46% |

*3.4. Factors which influence to the change of electricity demand*

The result of the partial adjustment model is presented in Table 7. In the period from spring to early summer, the number of the residential decreases the electricity, and the number of work correlates the demand. Late August and September are hot times in Japan. Staying at home continued during this period, and as the number of residences increased, the demand for electricity increased.

From October to the first half of December, demand is increasing even if the number of residences increases from autumn to winter and the number of works increases. The second half of December to the first half of January is a holiday in Japan. At this time, as in the previous period, demand is increasing regardless of whether the residence increases or the work increases.

During the cold season from late January to February, but from the beginning of the year, demand increases as more work is done.

March is a busy time to work at the end of the year in Japan. Demand increased regardless of whether the number of residences increased or the number of works increased. Conversely, as Retail increased, demand decreased.

**Table 7.** Estimation Results of the Partial Adjustment Model

| Periods | | $\ln Resi_{area,t}$ | $\ln Wrk_{area,t}$ | $\ln Grcy_{area,t}$ | $\ln Retl_{area,t}$ | $\ln ELE_{area,t-1}$ |
| --- | --- | --- | --- | --- | --- | --- |
| 1st April | 31st Jul | -0.1451* | 2.0061* | 0.8809* | -1.1772* | -0.1772* |
| 1st Aug | 15th Sep | 0.2428* | 0.2071 | 0.1417* | 0.0632 | 1.0632* |
| 16th Sep | 15th Dec | 0.1206* | -0.3099 | -0.179 | 0.2596 | 1.2596* |
| 16th Dec | 15th Jan | 0.4031* | 1.8525* | -0.3689 | 0.0978 | 1.0978* |
| 16th Jan | 31st Mar | 0.0059 | 0.2387* | -0.6671* | 0.2441 | 1.2441* |

\* Statistical threshold of z-value: 95%

## 4. Discussion

The significant decrease in electricity demand during the period from spring to early summer is thought to be due to the inactivity of people. The sharp increase in hot summer is thought to be due to the sharp increase in electricity demand at home because people stay at home. On the other hand, we cannot see statistical changes exist from late September to early December. It is pointed out that restrictions are relaxed during this period and people's lives are closer to normal. From the latter half of December to the first half of January, it increased again. During this period, the infection was spreading again, and many people stayed at home, which is thought to be due to an increase in household demand. The period from late January to March shows statistically insignificant as the demand shows in autumn.

*4.1. Understanding from the perspective of voltage level*

To examine the explanation above, we analyzed the demand for sectors. Electricity and Gas Market Surveillance Commission (2021) provides the monthly electricity demand for voltages (Extra high, High, Lighting, Low, Other voltage) which is reported by the electric power companies. The demand for each month in this report is defined that the difference of the accumulated demand in the previous month and the accumulated demand in this month, so the demand of this report differs from the electricity demand which is used in our analysis.

Figure 3 shows that the change of electricity demand in 2020-2021 from the demand in the average of 2018-2019 for the class of electricity voltage. The extra-high voltage, which is used in large factories, large business buildings, and shopping centers, and department stores, decreases in all months. The decreasing volume in April-May is the largest. High voltage, which is used in the mid-sized factories, buildings, and others, is the most decreasing in April-May. After August-September, the decreasing volume is smaller than the former of the period. The lighting voltage which is used for households increases in almost all period except for June-July and July-August.

In August-September and December-January, the total demand change is positive, while extra-high-voltage is decreased. In summer and winter time, the demand for household exceeds the business demand because household need a lot of electricity for air conditioning. In spring and autumn, the increase of the household does not overcome the decrease of business demand, in spite of the large and mid-sized buildings' closure.

In the period when the infection spread rapidly from spring to summer, it decreases electricity demand that people stay at home not using much electricity. On the other hand, in hot summers and cold winters, electricity consumption increased at home, and the relaxing behavioral restrictions might increase business demand for electricity.

**Figure 3.** Monthly Electricity demand for each voltage

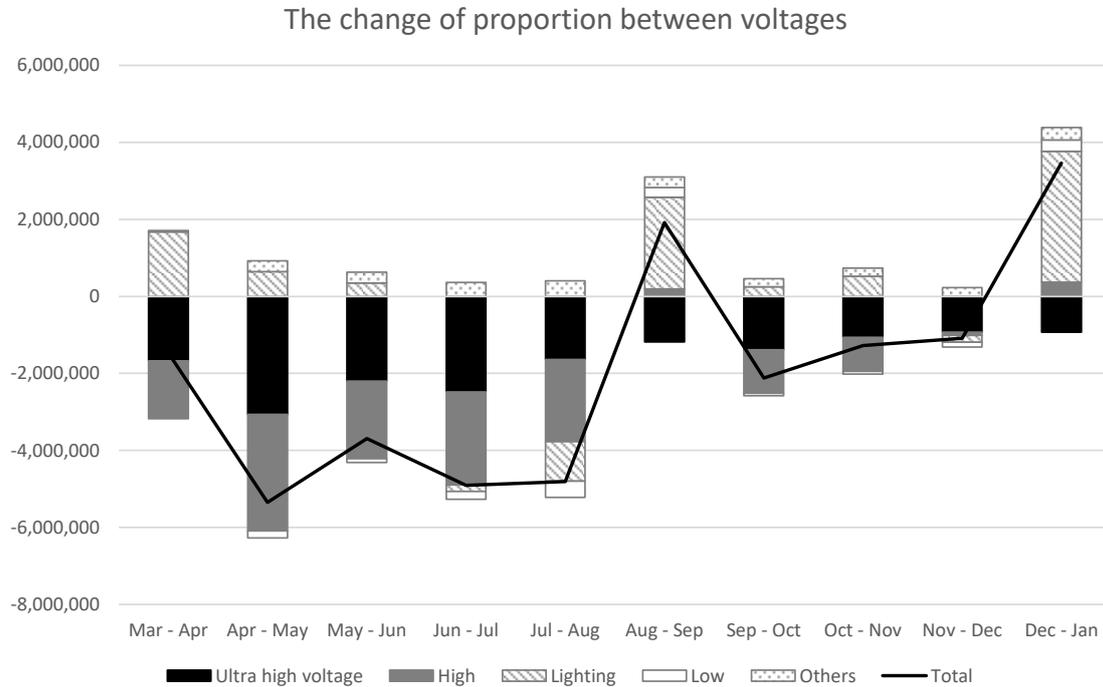

## 4.2. Policy implications

Policy makers may think that people's behavioral restrictions may be the factor which pushes down electricity demand, but a crisis of serious power shortages due to higher-than-normal demand in the period from mid-December 2020 to early January 2021. Although the direct cause of the crisis was record cold waves, generation troubles, and fuel procurement failures, it seems important to understand how much power demand actually increases or decreases. Electricity demand is estimated from the expected temperature according to the formula estimated from the accumulated data especially in Japan. It may not be possible to estimate using the formulas accumulated so far, when a serious event occurs. The methods and results presented in this study may help policy makers to expect the change rate of electricity demand, and the feature of our methods are that they need time-series data which policy makers have accumulated.

## 5. Conclusion

We conclude as follows. Electricity demand increases in the summer and winter even if severe events occur because the increase of demand for household sector may exceed the decrease of business demand, while the decrease of electricity demand for business sector may exceed the increase of the demand for household sector in the spring. The electricity demand in April and May decreases as the number of infected increases, but electricity demand in summer and winter tends to increase as the infected increases.

Now, we discuss the possible extensions of this study. We can straightforwardly extend this study to the supply side of electricity. As pointed out in this study, electricity demand rapidly decreases in spring, while the demand increases in summer and winter season, especially, the demand at peak time in these seasons increases keenly. This change would have impact on the electricity generation and electricity transmission sectors, especially the change of the peak hour will change the demand of renewable energy.

Appendix

**Table A1.** Statistical tests for homogeneity of variance

Levene's Test for Homogeneity of Variance (center = mean)

|  | Df | F value | Pr(>F) |
|---|---|---|---|
| group | 4 | 19.504 | 1.534e-14 *** |

Signif. codes:  0 '***' 0.001 '**' 0.01 '*' 0.05 '.' 0.1 ' ' 1

TukeyHSD(aov(temperature ~ seasons))

|  |  | diff | lwr | upr | p.adj |
|---|---|---|---|---|---|
| Season1 | Season2 | 7.82 | 5.85 | 9.79 | 0.00E+00 |
|  | Season3 | -4.23 | -5.81 | -2.66 | 0.00E+00 |
|  | Season4 | -15.44 | -17.73 | -13.16 | 0.00E+00 |
|  | Season5 | -11.21 | -12.88 | -9.55 | 0.00E+00 |
| Season2 | Season3 | -12.05 | -14.11 | -10.00 | 0.00E+00 |
|  | Season4 | -23.27 | -25.91 | -20.62 | 0.00E+00 |
|  | Season5 | -19.04 | -21.17 | -16.91 | 0.00E+00 |
| Season3 | Season4 | -11.21 | -13.58 | -8.84 | 0.00E+00 |
|  | Season5 | -6.98 | -8.76 | -5.21 | 0.00E+00 |
| Season4 | Season5 | 4.23 | 1.80 | 6.66 | 2.61E-05 |

# References

Abu-Rayash, A., & Dincer, I., 2020. Analysis of the electricity demand trends amidst the COVID-19 coronavirus pandemic. Energy Research and Social Science, 68(June), 101682. https://doi.org/10.1016/j.erss.2020.101682

Agdas, D., & Barooah, P., 2020. Impact of the COVID-19 Pandemic on the U.S. Electricity Demand and Supply: An Early View from Data. IEEE Access, 8, 151523–151534.

Alasali, F., Nusair, K., Alhmoud, L., & Zarour, E., 2021. Impact of the covid-19 pandemic on electricity demand and load forecasting. Sustainability (Switzerland), 13(3), 1–22.

Bahmanyar, A., Estebsari, A., & Ernst, D., 2020. The impact of different COVID-19 containment measures on electricity consumption in Europe. Energy Research and Social Science, 68(June), 101683.

Brodersen, K.,H., Gallusser, F., Koehler, J., Remy, N., and Scott, S.L., 2015. Inferring causal impact using bayesian structural time-series models. The Annals of Applied Statistics, 9(1), 247–274.

Buechler, E., Powell, S., Sun, T., Zanocco, C., Astier, N., Bolorinos, J., Flora, J., Boudet, H., Rajagopal, R., 2020. Power and the pandemic: Exploring global changes in electricity demand during COVID-19. ArXiv.

Carvalho, M., Bandeira de Mello Delgado, D., de Lima, K. M., de Camargo Cancela, M., dos Siqueira, C. A., de Souza, D. L. B., 2021. Effects of the COVID-19 pandemic on the Brazilian electricity consumption patterns. International Journal of Energy Research, 45(2), 3358–3364.

Chubu Electric Power Grid Company, Inc., 2021. https://powergrid.chuden.co.jp/denkiyoho/

Chugoku Electric Power Transmission & Distribution Company, Inc. (2021) https://www.energia.co.jp/nw/jukyuu/download.html

Electricity and Gas Market Surveillance Commission, 2021. "Denryoku torihiki no joukyou" in Japanese (Results of electricity demand). https://www.emsc.meti.go.jp/info/business/report/results.html

Elavarasan, R. M., Shafiullah, G.M., Raju, K., Mudgal, V., Arif, M.T., Jamal, T., Subramanian, S., Sriraja Balaguru, V.S., Reddy, K.S., Subramaniam, U., 2020. COVID-19: Impact analysis and recommendations for power sector operation, Appl. Energy. 279 115739. https://doi.org/10.1016/j.apenergy.2020.115739.

García, S., Parejo, A., Personal, E., Ignacio Guerrero, J., Biscarri, F., & León, C., 2021. A retrospective analysis of the impact of the COVID-19 restrictions on energy consumption at a disaggregated level. Applied Energy, 287(January), 116547.

Google, 2021. Community Mobility Report. https://www.google.com/covid19/mobility/

Hokkaido Electric Power Network, Inc., 2021. http://denkiyoho.hepco.co.jp/area_download.html

Hokuriku Electric Power Transmission & Distribution Company, 2021. http://www.rikuden.co.jp/nw/denki-yoho/#download

Hsiao, C., 2003. Analysis of Panel Data, Edition 2, Cambridge University Press, UK.

Kansai Transmission and Distribution, Inc., 2021. https://www.kansai-td.co.jp/denkiyoho/download/index.html

Kirli, D., Parzen, M., & Kiprakis, A. (2021). Impact of the COVID-19 Lockdown on the Electricity System of Great Britain: A Study on Energy Demand, Generation, Pricing and Grid Stability. Energies, 14(3), 635.

Kyushu Electric Power Transmission and Distribution Company, Inc., 2021. https://www.kyuden.co.jp/td_power_usages/pc.html

Märkle-Huß, J., Feuerriegela, S., Neumanna, D., 2018. Contract durations in the electricity market: Causal impact of 15 min trading on the EPEX SPOT market. Energy Economics, 69, 367–378.

Ruan, G., Wu, J., Zhong, H., Xia, Q., & Xie, L., 2021. Quantitative assessment of U.S. bulk power systems and market operations during the COVID-19 pandemic. Applied Energy, 286(August 2020), 116354.

Santiago, I., Moreno-Munoz, A., Quintero-Jiménez, P., Garcia-Torres, F., & Gonzalez-Redondo, M. J., 2021. Electricity demand during pandemic times: The case of the COVID-19 in Spain. Energy Policy, 148(September 2020).

Shikoku Electric Power Transmission & Distribution Company, Inc., 2021. https://www.yonden.co.jp/nw/denkiyoho/download.html

Tanishita, M., 2009. Estimation of regional price elasticities of household's electricity demand (in Japanese),. Journal of Japan Society of Energy and Resources 30: 1–7.

TEPCO Power Grid, 2021. https://www.tepco.co.jp/forecast/html/download-j.html

Tohoku Electric Power Company, Inc., 2021. https://setsuden.nw.tohoku-epco.co.jp/download.html

The Okinawa Electric Power Company, Inc., 2021. https://www.okiden.co.jp/denki2/dl/

Wakashiro, Y., 2020. Impact of COVID-19 on Electricity Demand in Japan. mimeo

Wakashiro, Y., 2019. Estimating price elasticity of demand for electricity: The case of Japanese manufacturing industry. International Journal of Economics Policy Studies. 13, 173–191.